# Itinerant Ferromagnetism in layered crystals LaCoO$X$ ($X$ = P, As)

Hiroshi Yanagi[1,*], Ryuto Kawamura[1], Toshio Kamiya[1,2], Yoichi Kamihara[2], Masahiro Hirano[2,3]
Tetsuya Nakamura[4], Hitoshi Osawa[4], and Hideo Hosono[1,2,3]

[1]Materials and Structures Laboratory, Tokyo Institute of Technology, 4259 Nagatsuta, Midori-ku, Yokohama 226-8503, Japan
[2]ERATO-SORST, Japan Science and Technology Agency (JST), in Frontier Research Center, Tokyo Institute of Technology, 4259 Nagatsuta, Midori-ku, Yokohama 226-8503, Japan
[3]Frontier Research Center, Tokyo Institute of Technology, 4259 Nagatsuta, Midori-ku, Yokohama 226-8503, Japan
[4]Japan Synchrotron Radiation Research Institute (JASRI), SPring-8, 1-1-1 Kouto, Sayo-cho, Sayo-gun, Hyogo 679-5198, Japan

The electronic and magnetic properties of cobalt-based layered oxypnictides, LaCoO$X$ ($X$ = P, As), are investigated. LaCoOP and LaCoOAs show metallic type conduction with room temperature resistivities of ~2×10$^{-4}$ Ω·cm, and the Fermi edge is observed by hard x-ray photoelectron spectroscopy. Ferromagnetic transitions occur at 43 K for LaCoOP and 66 K for LaCoOAs with spontaneous magnetic moments of 0.33 $\mu_B$ and 0.39 $\mu_B$ extrapolated to 0 K, respectively. Above the transition temperatures, the magnetic susceptibility exhibits a large temperature dependence. Provided that this temperature dependence follows the Curie-Weiss law, enhanced magnetic moment values of ~2.9 $\mu_B$ for LaCoOP and ~1.3 $\mu_B$ for LaCoOAs are obtained. X-ray magnetic circular dichroism (XMCD) is observed at the Co $L_{2,3}$-edge, but not at the other edges. The calculated electronic structure shows a spin polarized ground state with a magnetic moment of 0.52 $\mu_B$, ~95% of which is localized on the Co ions. These results indicate that LaCoO$X$ are itinerant ferromagnets and suggest that their magnetic properties are governed by spin fluctuation.

PACS No.:
75.50.Cc      (Other ferromagnetic metals and alloys)
74.25.Ha      (Magnetic properties)
74.25.Fy      (Transport properties (electric and thermal conductivity, thermoelectric effects, etc.))

---

Footnotes:
([*]) Electronic mail: yanagi@lucid.msl.titech.ac.jp

# I. Introduction

Correlated electron systems composed of *d* electrons in transition metal compounds have been studied intensively due to their distinct electronic and magnetic properties such as high transition temperature superconductivity, itinerant ferromagnetism, and giant magnetoresistance. Recently, our studies on 3*d*-transition-metal-based layered oxypnictides, La*MOX* (*M* = transition metal ion with a divalent formal charge, including Mn (electronic configuration of $3d^5$), Fe ($3d^6$), Co ($3d^7$), Ni ($3d^8$), and Zn ($3d^{10}$), *X* = P and As), as a new correlated electron system have led to the discovery of superconducting transitions in LaFeOP[1] and LaNiOP.[2] The crystal structure of La*MOX* is composed of alternating stacks of LaO and *MX* layers along the *c*-axis as shown in **Fig. 1**. In addition, our research on an isostructual compound LaCuO*Ch* (*Ch* = S, Se, Te), has revealed that the Cu*Ch* layer, which is the counterpart to the *MX* layer in La*MOX*, acts as a hole transport path, and that the holes are confined in the Cu*Ch* layer because it is sandwiched by wider energy gap LaO layers.[3] Due to the similarities between these compounds, we considered that the carrier electrons in La*MOX* are confined two-dimensionally in the *MX* layer, which causes strong electron-electron interactions. La*MOX* is much attractive because a transition metal ion with magnetic 3*d* electrons exists in the conduction layer, and various magnetic phases are expected to emerge as the number of 3*d* electrons is altered by replacing the *M* ion species. The magnetic properties of LaCoO*X* have attracted attention because LaFeOP and LaNiOP, which have transition metal elements with even numbers of 3*d* electrons (6 for Fe and 8 for Ni), undergo superconducting transitions.[1,2,4] Magnetic orderings are suppressed in these superconductors presumably due to reduction of magnetic moments of Fe and Ni. On the other hand, the magnetic moment of the Co ion would not vanish completely due to the odd number of 3*d* electrons, anticipating an appearance of the magnetic spin ordering phase. In addition, recently we have reported that an isostructural compound LaFeOAs exhibits a superconducting transition at a fairly high temperature ~26 K (on-set transition temperature is ~32 K) by doping F ions to the O ion sites.[5] In this case, the F doping is expected to increase the number of Fe 3*d* electrons closer to that of the Co 3*d* electrons in LaCoO*X*, which makes the electronic structure around the Fermi level similar for the F-doped LaFeOAs and LaCoO*X*. Therefore, it is interesting to examine the magnetic and electrical properties of LaCoO*X* in comparison with the LaFeO*X* system.

In this paper, we report temperature dependent electrical conductivity and magnetization and demonstrating that LaCoO*X* (*X* = P, As) are itinerant ferromagnets with Curie temperatures of 43 K for LaCoOP and 66 K for LaCoOAs. X-ray magnetic circular dichroism (XMCD) measurements and first-principles calculations for the electronic structure reinforce the conclusion that LaCoO*X* show itinerant ferromagnetisms. Ferromagnetic spin fluctuation of the carrier electrons may govern the magnetic properties of these compounds.

# II. Experimental

We employed different synthesis processes from those reported by Zimmer et al. for LaCoOP[6] and Quebe et al. for LaCoOAs.[7] That is, LaCoOP was synthesized by solid-state reactions of a stoichiometric mixture of LaP and CoO. First LaP was prepared by a solid-state reaction of La (Shinetsu, 99.5%) with P (Rare Metallic, 99.999%) at 700 °C for 10 h in evacuated silica tubes. Commercial CoO powders (Rare Metallic, 99.99%) were dehydrated at 1000 °C for 10 h in Ar. Then, LaCoOP was synthesized by heating a stoichiometric mixture of LaP and CoO powders in a silica tube at 1250 °C for 20 h. On the other hand, to synthesize LaCoOAs, LaAs was prepared by solid-state reactions between La and As (Kojundo Chemical Lab. 99.99%) at 700 °C for 5 h. Then a mixture of LaAs and dehydrated CoO powders was heated at 1100 °C for 20 h in evacuated silica tubes. The silica tubes were filled with high purity Ar gas (99.999%) at atmospheric pressure and room temperature to prevent implosion of the tube during the heating process. The colors of the purest LaCoOP and LaCoOAs powders were dark gray.

Phase purities and crystal structures of the synthesized powders were examined by powder x-ray diffraction (XRD, D8 ADVANCE, Bruker AXS) using Cu $K\alpha$ radiation from a rotating anode (45 kV × 360 mA) in a $2\theta$ range of 5 – 140° with a scanning speed of 0.64°/min. Crystal structures were refined by the Rietveld method using TOPAS 3 software[8] (Bruker AXS) based on a fundamental parameters approach. Electrical resistivities were measured by the four-probe method using a Physical Property Measurement System (PPMS, Quantum Design) in a temperature range from 2 to 320 K. For the measurements, rectangular shaped sintered pellets (apparent density ~60%) with plate Au electrodes deposited by sputtering were employed. Magnetization measurements were performed with a vibrating sample magnetometer (VSM, Quantum Design) from 2 to 300 K under magnetic fields ($H$) of 0 to 9 T. Magnetoresistance of LaCoOP was measured in the temperature range from 2 to 100 K under several $H$ between 0.1 – 6 T, which was applied perpendicular to the electric current.

Hard x-ray photoelectron spectroscopy (HXPS) measurements were performed with an excitation photon energy of 7936 eV in the BL29XU beamline at the Japan Synchrotron Radiation Research Institute (SPring-8).[9,10] The energies of the photoelectrons were analyzed using a Gammadata-Scienta R4000 electron spectrometer. X-ray absorption spectroscopy (XAS) and XMCD spectra were measured at La $M_{4,5}$-, Co $L_{2,3}$-, O $K$- and P $K$-edges in the BL25SU beamline at SPring-8.[11] XAS data were obtained by the total electron yield (TEY) method. On the other hand, the XMCD spectra were taken as the difference in the XAS spectra measured for different helicity x-rays under $H$ up to 1.9 T. Left- and right-handed circularly polarized radiations were generated by a twin helical undulator for the incident x-rays. Polycrystalline LaCoOP samples were fractured in the vacuum preparation chamber to obtain clean surfaces just before the measurements. All measurements were performed at 20 K in an ultra high vacuum of $\sim 10^{-8}$ Pa.

Spin-polarized electronic structures were calculated based on the linearized/augmented plane wave plus local orbitals (L/APW+lo) method with a density-functional theory (DFT) level using PBE96 functionals with the WIEN2k code.[12] Calculation results, which incorporated an electron correlation correction beyond GGA such as LDA/GGA+U are not shown because GGA+U calculations with the effective Coulomb parameters $U$ - $J$ varied from 0 to 7 eV gave essentially the same electronic structure (a metallic electronic structure with a small ferromagnetic polarization), which is reasonable because LaCoOP is a metal and the electrons are highly delocalized. Structural relaxation calculations were also performed, and the relaxed structures were compared to those obtained by Rietveld analyses. For the structure relaxation calculations, the plane wave and projector augmented wave (PAW) method[13, 14] with PBE96 functionals was employed using a VASP code.[15]

### III. Results and Discussion

**Figure 2(a)** shows the powder XRD pattern of LaCoOP. The simulated pattern by the Rietveld method agrees well with the observed one and no extra peak is observed, confirming that the sample is single-phase. On the other hand, major diffraction peaks in the XRD pattern of LaCoOAs in **Fig. 2(b)** are identified as those of the LaCoOAs phase, but weak peaks due to impurity phases, CoAs and $La_2O_3$, are detected. Three-phases Rietveld analysis estimates the impurity phases of CoAs and $La_2O_3$ to be ~12 mol% and ~1 mol%, respectively. It should be noted that $La_2O_3$ is a non-magnetic insulator and CoAs does not show magnetic ordering in the temperature range from 4.2 to 300 K.[16] Therefore, we neglected the effects of the impurities in the analysis of magnetism. **Table I** summarizes the structural parameters refined by Rietveld analyses, which are consistent with those from the structure relaxation calculations (also shown in **Table I**). That is, the DFT calculations reproduce the experimental structural parameters with errors less than 1.2% both for LaCoOP and LaCoOAs.

**Figure 3** shows electrical resistivities ($\rho$) of LaCoOP and LaCoOAs as a function of $T$ from 2 to 320 K. The $\rho$ values of LaCoOP and LaCoOAs at 300 K are $2.5 \times 10^{-4}$ Ω·cm and $2.3 \times 10^{-4}$ Ω·cm, respectively, which are one order of magnitude lower than those of the isostructural metallic compounds LaNiOP ($1.2 \times 10^{-3}$ Ω·cm) and LaFeOP ($1.6 \times 10^{-3}$ Ω·cm).[1,2] $\rho$ decreases with temperature, exhibiting metallic type conduction.

**Figure 4(a)** shows the $T$ dependences of molar magnetizations, which increase sharply with decreasing the temperature below ~50 K for LaCoOP and ~70 K for LaCoOAs. These observations suggest that LaCoO$X$ undergo magnetic transitions around these temperatures, respectively, and the magnetic phases below the transition temperature exhibit the spontaneous magnetic moments. **Figures 4(b) and (c)** show the magnetic field ($H$) dependences of the magnetic moments ($M$) for LaCoOP and LaCoOAs, respectively, measured at 2 – 80 K. At temperatures above 50 K for LaCoOP and 70 K for LaCoOAs, the $M$-$H$ curves are nearly linear, indicating that the compounds are paramagnetic at these temperatures. However, the decrease in $T$ changes the straight lines to S shaped curves, suggesting the emergence of a spontaneous magnetic moment ($M_S$), which increases with further decreasing $T$. The $M$-$H$ curves showed very small, but finite hysteresis (the measured coercive magnetic field is 2.5 mT for LaCoOP), indicating they are ferromagnets. **Figure 4(d)** shows $M_S$ as a function of $T$ obtained from the $M^2$-$H/M$ curves (Arrott plots) as well as the $T$ dependences of inverse molar susceptibilities ($1/\chi_{mol}$). The Curie temperatures ($T_c$) estimated from the Arrott plots are 43 K for LaCoOP and 66 K for LaCoOAs. The normalized $M_S$-$T$ curves (**inset of (d)**) show similar dependences both for LaCoOP and LaCoOAs. The $M_{S0}$ values estimated by extrapolating the $M_S$-$T$ curves to $T = 0$ K are 0.33 $\mu_B$ per Co for LaCoOP and 0.39 $\mu_B$ per Co for LaCoOAs. The $1/\chi_{mol}$ - $T$ plots display linear behaviors and follow the Curie-Weiss law above 200 K. However, the effective magnetic moments ($M_{eff}$) estimated from the slopes of the Curie-Weiss plots are ~2.9 $\mu_B$ for LaCoOP and ~1.3 $\mu_B$ for LaCoOAs, both of which are significantly larger than those estimated from the extrapolation of the $M_S$-$T$ curves.

**Figure 5** shows temperature dependence of the magnetoresistance ($MR$) ratio for LaCoOP under several magnetic fields. The $MR$ ratio - $T$ curves show minimums around the Curie temperature, and the largest $MR$ ratio exceeds -25% at $H = 6$ T. These results, the Curie-Weiss behavior in the paramagnetic phase, the minimums in the $MR$ ratio around $T_c$, and the enhanced $M_{eff}$ value compared to the $M_{S0}$ value, strongly suggest that the spin fluctuation governs the magnetic properties of LaCoO$X$. The inconsistency between $M_{S0}$ and $M_{eff}$ would be explained by the spin fluctuation theory.[17] It indicates that the $M_{eff}$ does not contribute to a static magnetic moment but to a dynamic magnetic response, suggesting fluctuation of local spin moments. The negative $MR$ is also explained by spin fluctuation because fluctuation of spins causes larger spin scattering and larger resistivity while the applied magnetic field forces a more ordered spin structure, suppresses the spin fluctuation, and enhances electron transport. We like to point that there are other possible mechanisms for the observed ferromagnetic behavior such as spin density wave, which is reported for undoped LaFeOAs,[18] but further investigations will be required to make a conclusion. Here, we safely conclude that LaCoO$X$ are itinerant ferromagnets.

**Figure 6** shows a HXPS spectrum of LaCoOP around the Fermi energy at 20 K. The photoelectron intensity steeply drops at a zero binding energy (note that the zero binding energy corresponds to the Fermi level), and the edge width is nearly the same as that of the gold reference (the gray line), indicating that LaCoOP shows a Fermi edge, which further substantiates LaCoOP is metal. **Figures 7(a)** and **(b)** show the XAS and XMCD spectra of LaCoOP for the Co $L_{2,3}$-edge, respectively. The Co $L_{2,3}$-edge spectrum shows distinct XMCD signals, whereas other edges, including the La $M_{4,5}$-, O $K$- and P $K$-edges, do not exhibit noticeable XMCD signals. **Figure 7(c)** shows the $H$ dependence of the XMCD signal at the photon energy of 778 eV, where the absolute XMCD signal reaches the maximum. The curve is similar to the $M$-$H$ curve in **Fig. 4(b)**. These results suggest that the ferromagnetism originates dominantly from the Co 3$d$ electrons. The estimated magnetic moment of the Co using the XMCD sum rule is 0.14 $\mu_B$ with an accuracy of ~10%, which gives spin and orbital contributions of 0.12 $\mu_B$ and 0.02 $\mu_B$, respectively.[19-21] The estimated magnetic moment of 0.14 $\mu_B$ is smaller than the $M_S$ of 0.27 $\mu_B$ at 20 K (**Fig. 4(d)**). A possible reason would be that the observed XMCD signal was underestimated because XMCD is sensitive to a surface region and the fractured surface used for the XMCD measurements would have a magnetic dead layer.

**Figure 8** shows the total density of states (DOS) of LaCoOP around the Fermi energy calculated by the spin-polarized DFT. It shows that finite DOSs exist at the Fermi level, indicating that LaCoOP has a metallic electronic structure, which agrees with the metallic conduction and the Fermi edge structure in the HXPS spectrum. In addition, the electronic structure of LaCoOP is spin polarized at the ground state. The calculated magnetic moment is 0.52 $\mu_B$, which agrees reasonably well with the experimental values of 0.33 $\mu_B$ estimated from the $M_S$-$T$ curve. The DFT result also shows that most of the magnetic moments localize in the Co ions and the contribution from the P ions is less than 5%. This is consistent with no appearance of XMCD signal at the P $K$-edge. **Figures 9(a)** and **(b)** respectively show the up and down spin projected densities of states (PDOSs) of P 3$p$, Co 4$s$, 4$p$ and 3$d$ orbitals in LaCoOP, where the PDOSs of P 3$p$ and Co 3$d$ are further decomposed into the irreducible representations of the atomic orbitals in the Muffin-Tin spheres used for the L/APW+lo calculations. These show that P 3$p$, Co 3$d$, 4$s$, and 4$p$ orbitals form highly hybridized orbitals in the valence band and around the Fermi level. All these orbitals, especially Co 3$d_{x^2-y^2}$, have large contributions to the Fermi level. **Figure 9(c)** illustrates a simplified energy diagram of the irreducible representations of the Co $d$ orbitals deduced from the PDOSs. It shows that the $d$ orbitals are primary classified into two groups: the lower ($d_{xy}$) & ($d_{xz}$, $d_{yz}$) group and the upper ($d_{x^2-y^2}$) & ($d_{z^2}$) group. Moreover, the figure shows that the upper ($d_{x^2-y^2}$) & ($d_{z^2}$) states mainly form the Fermi level and induce the magnetic moment.

### IV. Summary

Magnetic and electrical measurements indicated that LaCoO$X$ are itinerant ferromagnets with transition temperatures and spontaneous magnetic moments of 43 K and 0.33 $\mu_B$ for LaCoOP and 66 K and 0.39 $\mu_B$ for LaCoOAs, respectively. XMCD measurements and DFT calculations further supports this conclusion, revealing that hybridized orbitals of Co 3$d$ and P 3$p$ mainly form the Fermi level and that the Co 3$d$ is responsible mostly for the magnetic moments.

### Acknowledgments

We are indebted to Drs. Keisuke Kobayashi, Eiji Ikenaga, Jung Jin Kim, Masaaki Kobata, Sigenori Ueda (JASRI, SPring-8) for their support with the HXPS measurements and to Drs. Hidenori Hiramatsu and Kentaro Kayanuma (ERATO-SORST, JST) for their help with the measurements and their valuable discussions. The XAS and XMCD measurements were performed at SPring-8 with the approval of JASRI as a Nanotechnology Support Project of MEXT. (Proposal No. 2006A1665/BL25SU)

**TABLE 1.** Crystal structures of LaCoO$X$ refined by Rietveld analyses and obtained by structure relaxation calculations by DFT. Those reported by Zimmer et al.[6] and by Quebe et al.[7] are shown for comparison (note that Ref. 7 reports only the lattice parameters).

|  |  | $a$ (nm) | $c$ (nm) | $R_p$ (%) | $R_{wp}$ (%) | $R_e$ (%) | $S$ |
|---|---|---|---|---|---|---|---|
| LaCoOP | Rietveld | 0.39681(9) | 0.83779(1) | 4.08 | 5.28 | 3.69 | 1.43 |
|  | Ref. **6** | 0.39678(9) | 0.8379(3) | 1.8 | 2.3 | - | - |
|  | DFT | 0.3940 | 0.8376 | - | - | - | - |
| LaCoOAs | Rietveld | 0.40526(1) | 0.84620(4) | 15.26 | 21.56 | 16.71 | 1.29 |
|  | Ref. **7** | 0.4054(1) | 0.8472(3) | - | - | - | - |
|  | DFT | 0.4039 | 0.8421 | - | - | - | - |

|  | $x$ | $y$ | $z$ | $z$ (Ref. **6**) | $z$ (DFT) |
|---|---|---|---|---|---|
| La | 1/4 | 1/4 | 0.1509(5) | 0.155(1) | 0.1526 |
| Co | 3/4 | 1/4 | 1/2 | 1/2 | 1/2 |
| O | 3/4 | 1/4 | 0 | 0 | 0 |
| P | 1/4 | 1/4 | 0.6321(9) | 0.617(6) | 0.6273 |

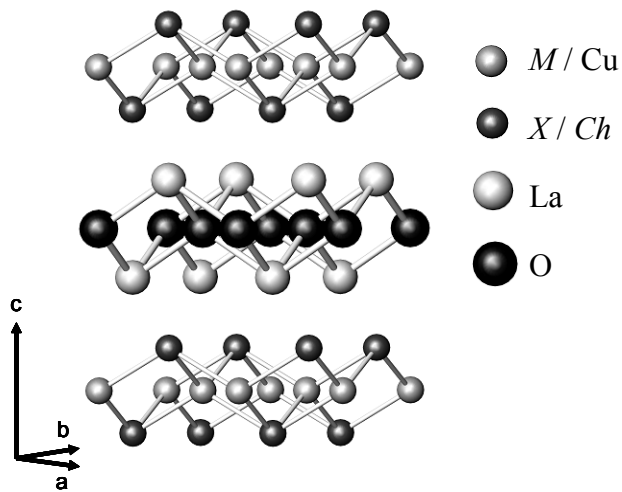

**FIG. 1.** Crystal structure of La*MOX* (*M* = Mn, Fe, Co, Ni, Zn, *X* = P, As) and LaCuO*Ch* (*Ch* = S, Se, Te), which have layered crystal structures composed of LaO and *MX*/Cu*Ch* layers.

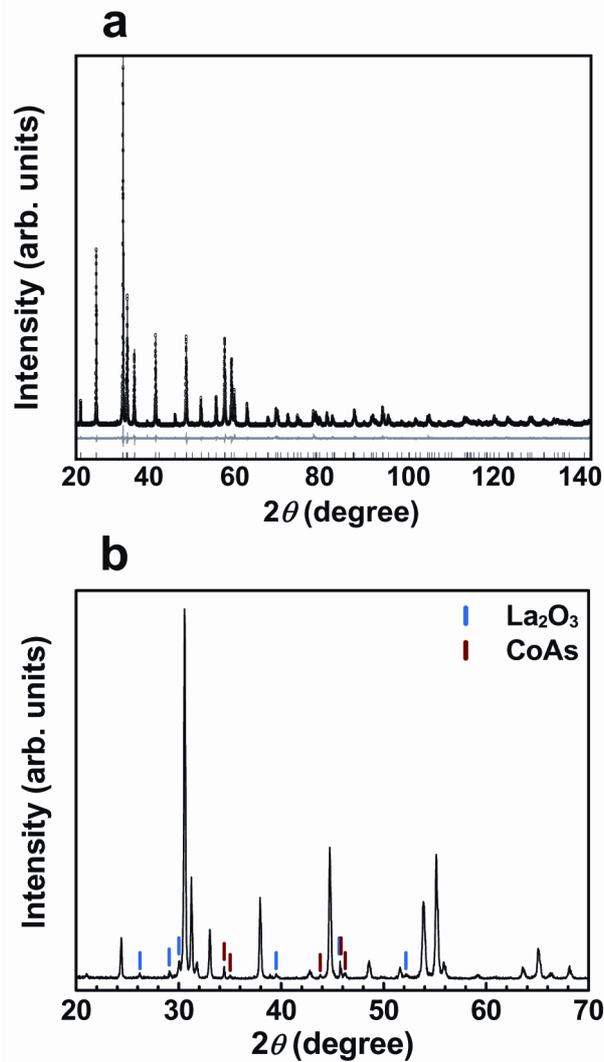

**FIG. 2.** (a) Powder XRD patterns of the LaCoOP sample and the result of Rietveld analysis. (Upper row) Observed pattern (circles) and simulated pattern obtained by Rietveld analysis (black line). (Middle row) Difference profile between the observed and simulated patterns. (Bottom row) Positions of Bragg reflections from LaCoOP. (b) Observed powder XRD pattern of the LaCoOAs sample. Red and blue vertical bars denote peak positions of CoAs and $La_2O_3$, respectively. Other peaks are explained by the LaCoOAs phase.

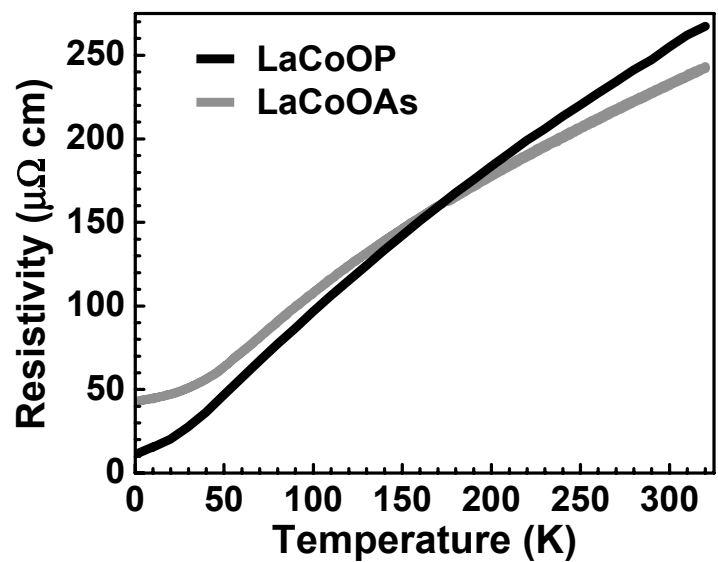

**FIG. 3.** Electrical resistivities of LaCoOP (black line) and LaCoOAs (gray line) sintered disks as a function of temperature.

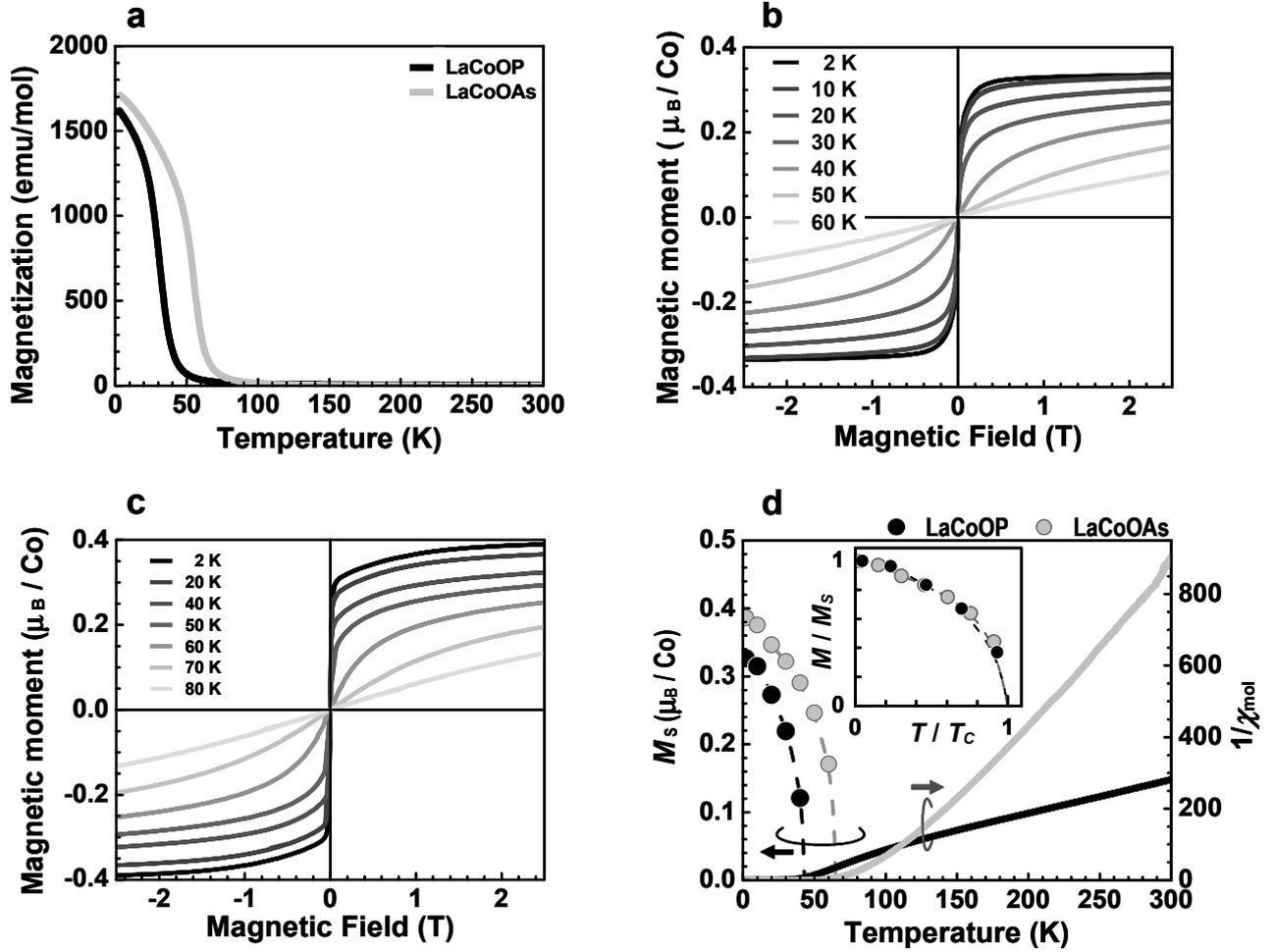

**FIG. 4.** (a) Temperature dependences of molar magnetization of LaCoOP (black line) and LaCoOAs (gray line) at $H = 0.1$ T. Magnetic field dependences of the magnetic moment of (b) LaCoOP and (c) LaCoOAs at various temperatures. (d) Spontaneous magnetic moment ($M_S$) and $1/\chi_{mol}$ of LaCoOP (black solid circles and black dashed line, and black solid line, respectively) and LaCoOAs (gray solid circles and gray lines) as functions of temperature. Inset shows the normalized $M_S$-$T$ curves.

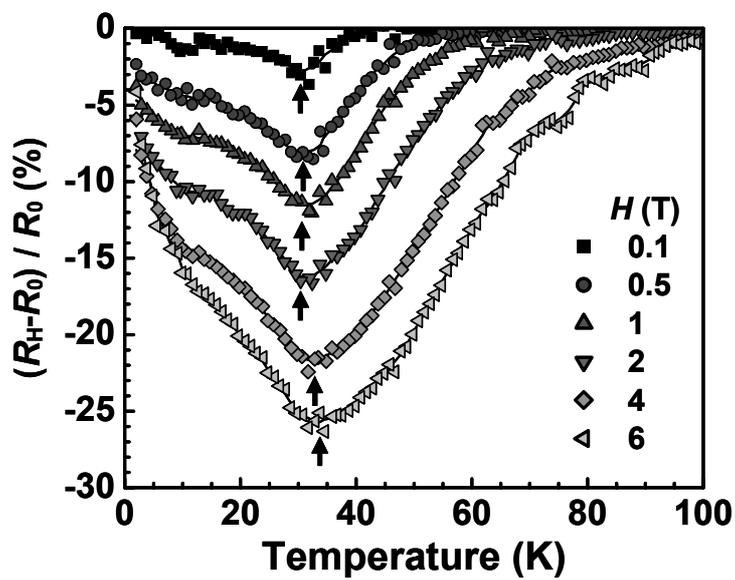

**FIG. 5.** Temperature dependences of the magnetoresistance ratio of LaCoOP at magnetic fields from 0.1 to 6 T. $R_0$ and $R_H$ denote resistances at a zero magnetic field and at an applied magnetic field of $H$, respectively.

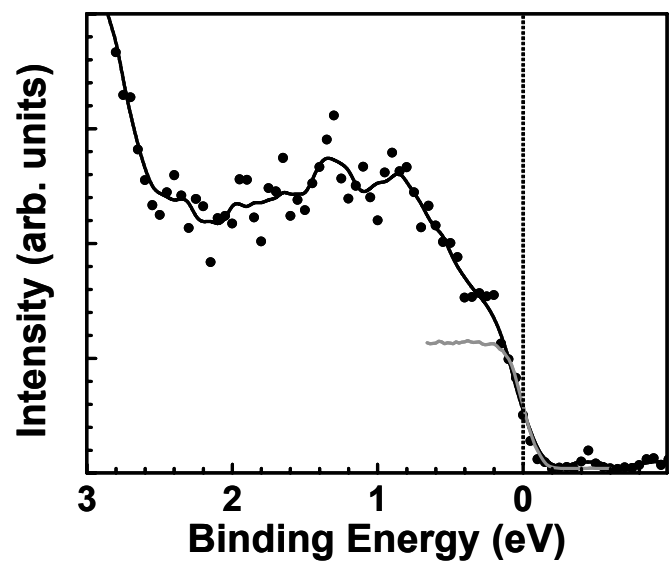

**FIG. 6.** Fermi edge region HXPS spectrum of LaCoOP. Circles show measured data and black line shows a smoothed curve. Gray line shows the Fermi edge spectrum of an Au reference.

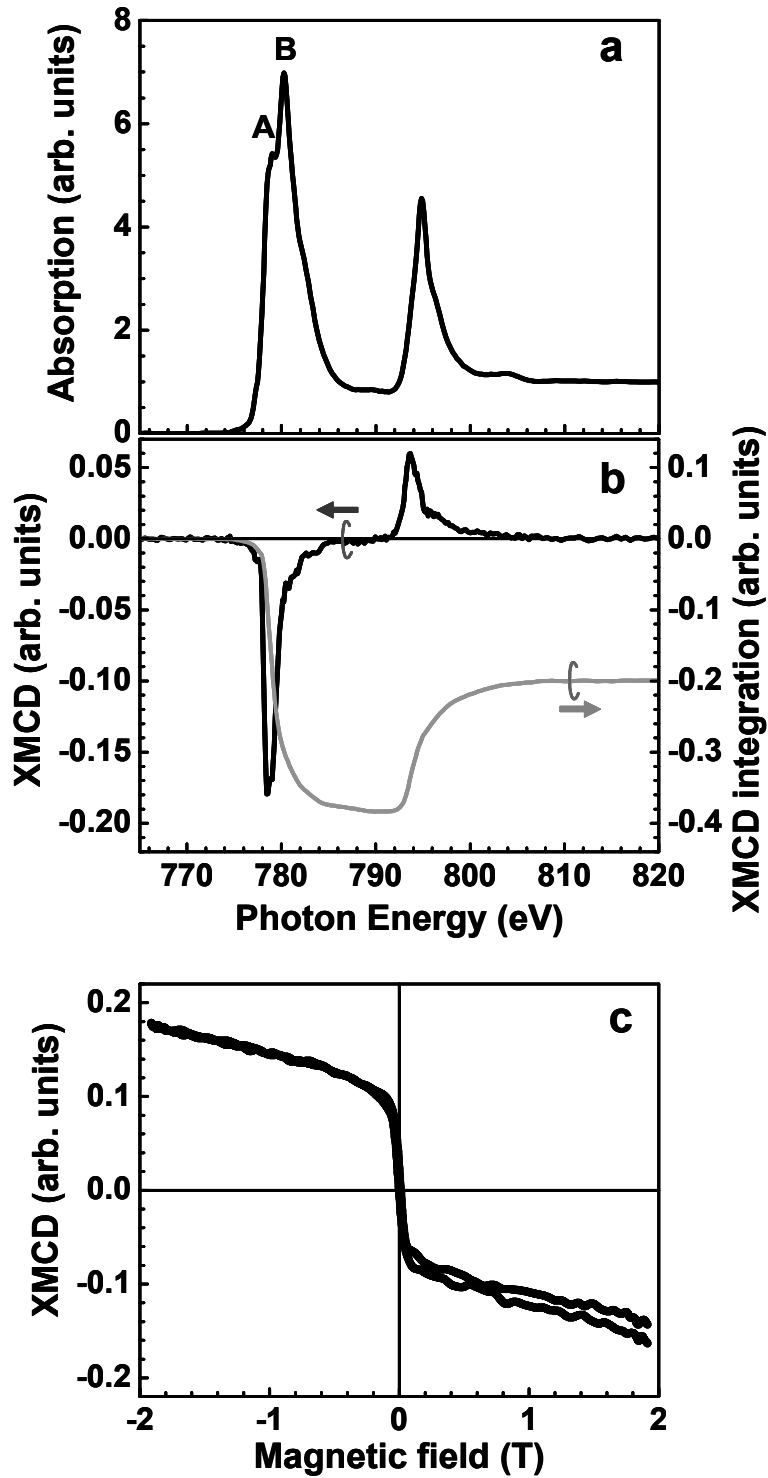

**FIG. 7.** (a) Co $L_{2,3}$-edge XAS spectra of LaCoOP. (b) Co $L_{2,3}$-edge XMCD spectrum (black line) and integrated XMCD curves (gray line). (c) Magnetic field dependence of the XMCD signal at the photon energy of 778 eV. All the measurements were performed at 20 K

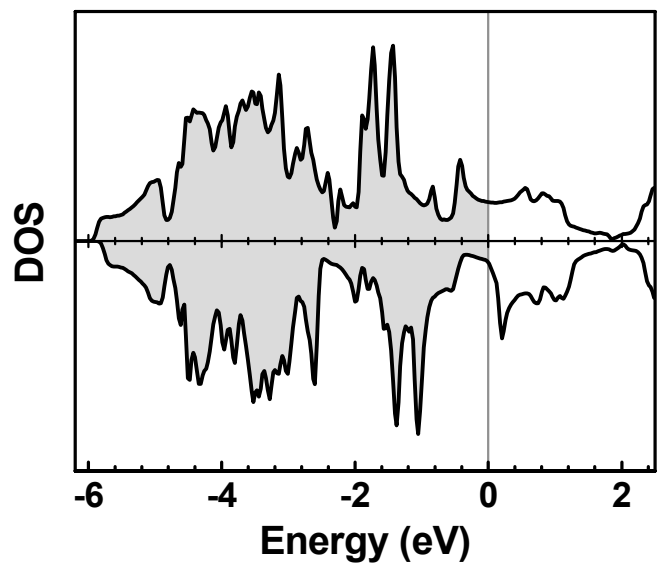

**FIG. 8.** Calculated total density of states (DOS) of LaCoOP with $U - J = 0$ near the Fermi level. Top and bottom panels show up and down spin DOSs, respectively. The energy is measured from the Fermi level and the shadowed area indicates occupied states.

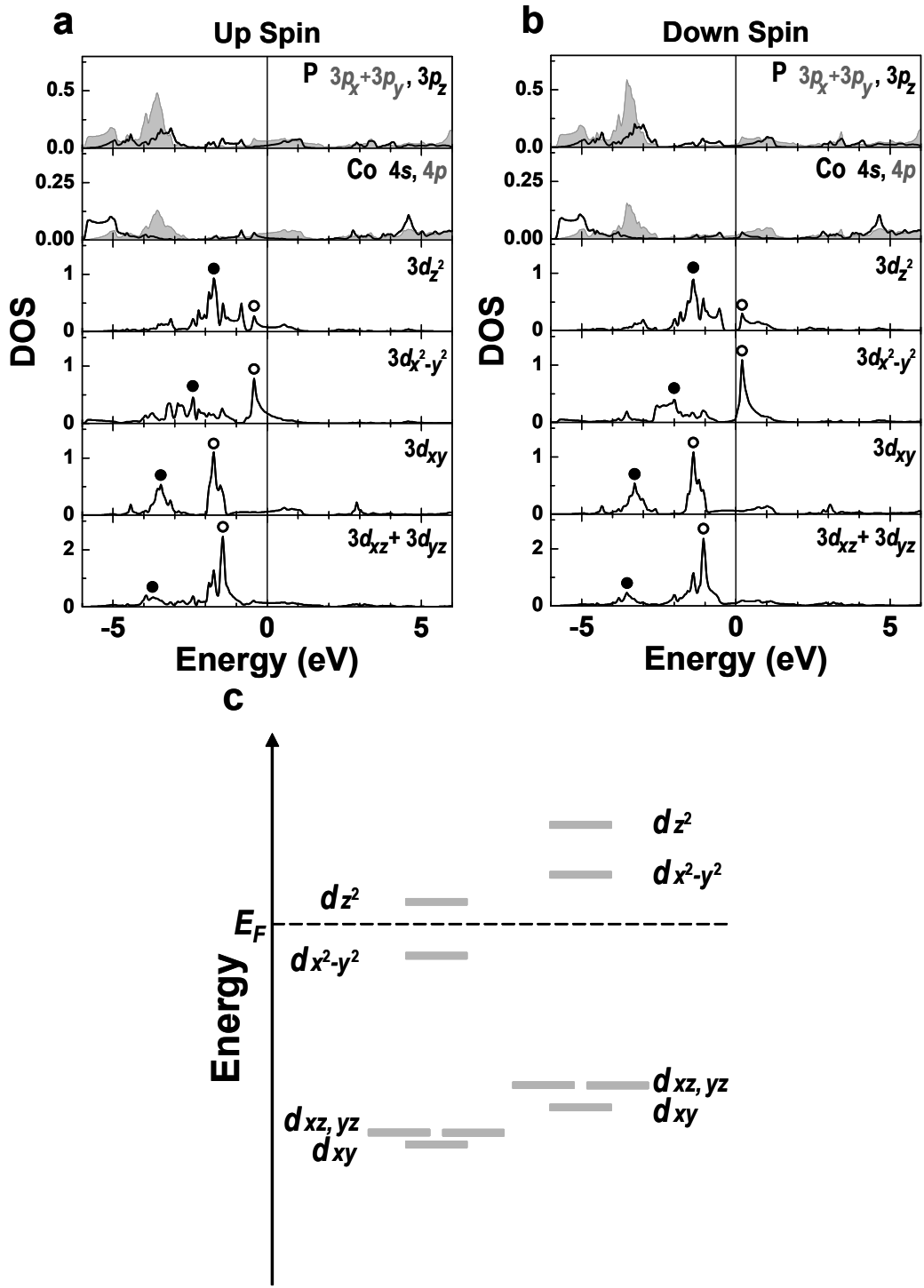

**FIG. 9.** PDOSs of P $3p_x + 3p_y$, $3p_z$ and Co $4s$, $4p$, $3d_{z^2}$, $3d_{x^2-y^2}$, $3d_{xy}$, $3d_{xz} + 3d_{yz}$ for (a) up and (b) down spins. Solid and open circles indicate major peaks, which correspond to bonding and anti-bonding states, respectively. (c) Simplified energy diagram of Co $3d$ levels, designated by the irreducible representations of the Co site. Energy levels are from the representative peaks in (a) and (b). Dashed line indicates the Fermi level.

16